\pdfoutput=1

\documentclass[12pt,a4paper]{article}

\usepackage{ifthen} 
\usepackage{rotating}
\usepackage{dashrule}
\usepackage{array} 
\usepackage{dcolumn}
\usepackage{comment}
\usepackage{xcolor}

\usepackage{graphpap} 
\usepackage{rotating} 
\usepackage{tikz}
\usepackage{xcolor}
\usepackage{longtable} 

\usepackage{lscape}
\usepackage{multirow} 
\usepackage{mathrsfs}
\usepackage{mathtools} 
\usetikzlibrary{patterns}
\usepackage{nicefrac} 
\usepackage{transparent}
\usepackage{afterpage} 
\usepackage{comment} 
\usepackage[normalem]{ulem}
\usepackage{cancel} 

\newboolean{pdflatex}
\setboolean{pdflatex}{true} 

\newboolean{articletitles}
\setboolean{articletitles}{true} 

\newboolean{uprightparticles}
\setboolean{uprightparticles}{True} 

\def\paperauthors{T.~Ovsiannikova on behalf of LHCb collaboration} 
\def\paperasciititle{Study of BsjpsipipiKK decays} 
\def\papertitle{Spectroscopy in beauty decays at the LHCb experiment} 
\def\paperkeywords{{High Energy Physics}, {LHCb}} 
\def\papercopyright{\the\year\ CERN for the benefit of the LHCb collaboration} 

\def\paperlicenceurl{https://creativecommons.org/licenses/by/4.0/}

\makeatletter
\g@addto@macro\bfseries{\boldmath}
\makeatother

\usepackage[top=1in, bottom=1.25in, left=1in, right=1in]{geometry}

\usepackage{tikz}

\usepackage{microtype}
\usepackage{lineno}  
\usepackage{xspace} 
\usepackage{caption} 

\usepackage{graphicx}  
\usepackage{color}
\usepackage{colortbl}
\graphicspath{{./figs/}} 
\DeclareGraphicsExtensions{.pdf,.PDF,png,.PNG}

\usepackage{amsmath} 
\usepackage{amssymb}
\usepackage{amsfonts}
\usepackage{upgreek} 

\newcommand*\patchAmsMathEnvironmentForLineno[1]{%
\expandafter\let\csname old#1\expandafter\endcsname\csname #1\endcsname
\expandafter\let\csname oldend#1\expandafter\endcsname\csname
end#1\endcsname
 \renewenvironment{#1}%
   {\linenomath\csname old#1\endcsname}%
   {\csname oldend#1\endcsname\endlinenomath}%
}
\newcommand*\patchBothAmsMathEnvironmentsForLineno[1]{%
  \patchAmsMathEnvironmentForLineno{#1}%
  \patchAmsMathEnvironmentForLineno{#1*}%
}
\AtBeginDocument{%
\patchBothAmsMathEnvironmentsForLineno{equation}%
\patchBothAmsMathEnvironmentsForLineno{align}%
\patchBothAmsMathEnvironmentsForLineno{flalign}%
\patchBothAmsMathEnvironmentsForLineno{alignat}%
\patchBothAmsMathEnvironmentsForLineno{gather}%
\patchBothAmsMathEnvironmentsForLineno{multline}%
\patchBothAmsMathEnvironmentsForLineno{eqnarray}%
}

\usepackage{hyperxmp}

\usepackage[pdftex,
            pdfauthor={\paperauthors},
            pdftitle={\paperasciititle},
            pdfkeywords={\paperkeywords},
            pdfcopyright={Copyright (C) \papercopyright},
            pdflicenseurl={\paperlicenceurl}]{hyperref}

\usepackage[colorinlistoftodos,textsize=scriptsize]{todonotes}

\usepackage[all]{hypcap} 

\usepackage{ifthen} 
\setboolean{uprightparticles}{True} 
\usepackage{xspace} 
\usepackage{upgreek}

\def\lhcb   {\mbox{LHCb}\xspace}

\def\MagUp {\mbox{\em Mag\kern -0.05em Up}\xspace}

\ifthenelse{\boolean{uprightparticles}}%
{

 \def\Pdelta      {\ensuremath{\updelta}\xspace}

 \def\Pmu         {\ensuremath{\upmu}\xspace}

 \def\Ppi         {\ensuremath{\uppi}\xspace}

 \def\Pphi        {\ensuremath{\upphi}\xspace}                 
                  
 \def\Pchi        {\ensuremath{\upchi}\xspace}                 
 \def\Ppsi        {\ensuremath{\uppsi}\xspace}

 \def\PDelta      {\ensuremath{\Delta}\xspace}                 
 \def\PXi         {\ensuremath{\Xi}\xspace}                 
 \def\PLambda     {\ensuremath{\Lambda}\xspace}                 
 \def\PSigma      {\ensuremath{\Sigma}\xspace}                 
 \def\POmega      {\ensuremath{\Omega}\xspace}                 
 \def\PUpsilon    {\ensuremath{\Upsilon}\xspace}

 \def\PB      {\ensuremath{\mathrm{B}}\xspace}                 
                  
 \def\PD      {\ensuremath{\mathrm{D}}\xspace}

 \def\PJ      {\ensuremath{\mathrm{J}}\xspace}                 
 \def\PK      {\ensuremath{\mathrm{K}}\xspace}

 \def\PX      {\ensuremath{\mathrm{X}}\xspace}

 \def\Pb      {\ensuremath{\mathrm{b}}\xspace}                 
 \def\Pc      {\ensuremath{\mathrm{c}}\xspace}

 \def\Pi      {\ensuremath{\mathrm{i}}\xspace}

 \def\Ps      {\ensuremath{\mathrm{s}}\xspace}

 \def\thebaroffset{0.0em}
}
{

 \def\Pdelta      {\ensuremath{\delta}\xspace}

 \def\Pmu         {\ensuremath{\mu}\xspace}

 \def\Ppi         {\ensuremath{\pi}\xspace}

 \def\Pphi        {\ensuremath{\phi}\xspace}                 
                  
 \def\Pchi        {\ensuremath{\chi}\xspace}                 
 \def\Ppsi        {\ensuremath{\psi}\xspace}                 
                  
 \mathchardef\PDelta="7101
 \mathchardef\PXi="7104
 \mathchardef\PLambda="7103
 \mathchardef\PSigma="7106
 \mathchardef\POmega="710A
 \mathchardef\PUpsilon="7107
                  
 \def\PB      {\ensuremath{B}\xspace}                 
                  
 \def\PD      {\ensuremath{D}\xspace}

 \def\PJ      {\ensuremath{J}\xspace}                 
 \def\PK      {\ensuremath{K}\xspace}

 \def\PX      {\ensuremath{X}\xspace}

 \def\Pb      {\ensuremath{b}\xspace}                 
 \def\Pc      {\ensuremath{c}\xspace}

 \def\Pi      {\ensuremath{i}\xspace}

 \def\Ps      {\ensuremath{s}\xspace}

 \def\thebaroffset{0.18em}
}
\newcommand{\offsetoverline}[2][\thebaroffset]{\kern #1\overline{\kern -#1 #2}}%

\makeatletter
\ifcase \@ptsize \relax
  \newcommand{\miniscule}{\@setfontsize\miniscule{4}{5}}
\or
  \newcommand{\miniscule}{\@setfontsize\miniscule{5}{6}}
\or
  \newcommand{\miniscule}{\@setfontsize\miniscule{5}{6}}
\fi
\makeatother

\DeclareRobustCommand{\optbar}[1]{\shortstack{{\miniscule (\rule[.5ex]{1.25em}{.18mm})}
  \\ [-.7ex] $#1$}}

\def\mup        {{\ensuremath{\Pmu^+}}\xspace}
\def\mun        {{\ensuremath{\Pmu^-}}\xspace} 

\def\squark    {{\ensuremath{\Ps}}\xspace}
\def\squarkbar {{\ensuremath{\overline \squark}}\xspace}

\def\cquark    {{\ensuremath{\Pc}}\xspace}
\def\cquarkbar {{\ensuremath{\overline \cquark}}\xspace}

\def\bquark    {{\ensuremath{\Pb}}\xspace}

\def\pion   {{\ensuremath{\Ppi}}\xspace}

\def\pip    {{\ensuremath{\pion^+}}\xspace}
\def\pim    {{\ensuremath{\pion^-}}\xspace}

\def\kaon    {{\ensuremath{\PK}}\xspace}
\def\Kbar    {{\ensuremath{\offsetoverline{\PK}}}\xspace}

\def\KorKbar {\kern \thebaroffset\optbar{\kern -\thebaroffset \PK}{}\xspace}

\def\Kp      {{\ensuremath{\kaon^+}}\xspace}
\def\Km      {{\ensuremath{\kaon^-}}\xspace}

\def\Kstarz  {{\ensuremath{\kaon^{*0}}}\xspace}
\def\Kstarzb {{\ensuremath{\Kbar{}^{*0}}}\xspace}

\def\DorDbar {\kern \thebaroffset\optbar{\kern -\thebaroffset \PD}\xspace}

\def\B       {{\ensuremath{\PB}}\xspace}

\def\BorBbar {\kern \thebaroffset\optbar{\kern -\thebaroffset \PB}\xspace}

\def\Bd      {{\ensuremath{\B^0}}\xspace}

\def\BdorBdbar {\kern \thebaroffset\optbar{\kern -\thebaroffset \Bd}\xspace}
\def\Bu      {{\ensuremath{\B^+}}\xspace}

\def\Bp      {{\ensuremath{\Bu}}\xspace}

\def\Bs      {{\ensuremath{\B^0_\squark}}\xspace}

\def\BsorBsbar {\kern \thebaroffset\optbar{\kern -\thebaroffset \Bs}\xspace}

\def\jpsi     {{\ensuremath{{\PJ\mskip -3mu/\mskip -2mu\Ppsi\mskip 2mu}}}\xspace}
\def\psitwos  {{\ensuremath{\Ppsi{(\rm{2S})}}}\xspace}

\def\chicone  {{\ensuremath{\Pchi_{\cquark 1}}}\xspace}
\def\chiconex {{\ensuremath{\Pchi_{\cquark 1}(3872)}}\xspace}

\def\Y#1S{\ensuremath{\PUpsilon{(#1S)}}\xspace}

\def\LorLbar     {\kern \thebaroffset\optbar{\kern -\thebaroffset \PLambda}\xspace}

\def\BF         {{\ensuremath{\mathcal{B}}}\xspace}
\def\BRN         {\BF}

\newcommand{\decay}[2]{\ensuremath{#1\!\to #2}\xspace} 

\def\to                 {\ensuremath{\rightarrow}\xspace}

\def\AT#1     {\ensuremath{A_{\mathrm{T}}^{#1}}\xspace}           

\def\C#1      {\ensuremath{\mathcal{C}_{#1}}\xspace}                       
\def\Cp#1     {\ensuremath{\mathcal{C}_{#1}^{'}}\xspace}                    
\def\Ceff#1   {\ensuremath{\mathcal{C}_{#1}^{\mathrm{(eff)}}}\xspace}        
\def\Cpeff#1  {\ensuremath{\mathcal{C}_{#1}^{'\mathrm{(eff)}}}\xspace}       
\def\Ope#1    {\ensuremath{\mathcal{O}_{#1}}\xspace}                       
\def\Opep#1   {\ensuremath{\mathcal{O}_{#1}^{'}}\xspace}                    

\newcommand{\aunit}[1]{\ensuremath{\text{\,#1}}}       

\newcommand{\tev}{\aunit{Te\kern -0.1em V}\xspace}
\newcommand{\gev}{\aunit{Ge\kern -0.1em V}\xspace}
\newcommand{\mev}{\aunit{Me\kern -0.1em V}\xspace}
\newcommand{\kev}{\aunit{ke\kern -0.1em V}\xspace}
\newcommand{\ev}{\aunit{e\kern -0.1em V}\xspace}
\newcommand{\mevc}{\ensuremath{\aunit{Me\kern -0.1em V\!/}c}\xspace}
\newcommand{\gevc}{\ensuremath{\aunit{Ge\kern -0.1em V\!/}c}\xspace}
\newcommand{\mevcc}{\ensuremath{\aunit{Me\kern -0.1em V\!/}c^2}\xspace}
\newcommand{\gevcc}{\ensuremath{\aunit{Ge\kern -0.1em V\!/}c^2}\xspace}

\def\gsim{{~\raise.15em\hbox{$>$}\kern-.85em
          \lower.35em\hbox{$\sim$}~}\xspace}
\def\lsim{{~\raise.15em\hbox{$<$}\kern-.85em
          \lower.35em\hbox{$\sim$}~}\xspace}

\def\sqs   {\ensuremath{\protect\sqrt{s}}\xspace}

\def\tell1  {TELL1\xspace}
\def\ukl1   {UKL1\xspace}

\newcommand{\kevcc}{\ensuremath{\aunit{ke\kern -0.1em V\!/}c^2}\xspace}

\def\chiconex  {\ensuremath{\chicone(3872)}\xspace}

\usepackage{boldline}
\usepackage{bm}

\usepackage{cite} 
\usepackage{mciteplus}

\usepackage{longtable} 

\newcolumntype{d}[1]{D{,}{\,\pm\,}{#1} }
\newcolumntype{f}[1]{D{,}{.}{#1} }

\usepackage{tikz}

\begin{document}

\renewcommand{\thefootnote}{\fnsymbol{footnote}}
\setcounter{footnote}{1}


\begin{titlepage}
\pagenumbering{roman}

\vspace*{-1.5cm}
\vspace*{1.5cm}
\noindent
\begin{tabular*}{\linewidth}{lc@{\extracolsep{\fill}}r@{\extracolsep{0pt}}}
\\
 & & \today \\ 
\end{tabular*}

\vspace*{3.2cm}

{\normalfont\bfseries\boldmath\huge
\begin{center}
  \papertitle 
\end{center}
}

\vspace*{0.2cm}

\begin{center}
Tatiana Ovsiannikova\footnote{\tt{E-mail:}\href{email:Tatiana.Ovsiakkinova@cern.ch}{\tt{Tatiana.Ovsiannikova@cern.ch}}} on behalf of the~LHCb collaboration 
\\{\it Institute for Theoretical and Experimental Physics, NRC Kurchatov Institute, B.~Cheremushkinskaya~25, Moscow, 117218, Russia.}

\end{center}

\vspace{2cm}

\begin{abstract}

The beauty hadron decays is unique laboratory to study charmonium and charmonium-like states,
such as the $\chiconex$ meson, other exotic states and the
tensor $D$\nobreakdash-wave $\Ppsi_2(3823)$ states. However  the nature of many exotic charmonium-like candidates are still unknown. 
The most recent LHCb results related to b\nobreakdash-hadron 
decays to charmonium states and obtained using large 
data samples collected during the Run~1 and Run~2 
periods are presented. 
This includes the most precise determination 
of the mass and width of  the $\chiconex$ state 
using the ~\mbox{$\decay{\Bp}{ \jpsi \pip \pim \Kp}$} decays, observation of  a~ resonant structure denoted as X(4740) 
in the $\jpsi \Pphi$ mass spectrum from 
$\decay{\Bs}{\jpsi\pip\pim\Kp\Km}$~decays
and the~precise measurement of the~\Bs~meson mass.
\end{abstract}

\vspace{\fill}

\begin{center}
  Presented at XXVII Cracow EPIPHANY Conference on future of particle physics
\end{center}



\end{titlepage}


\newpage
\setcounter{page}{2}
\mbox{~}
%

\cleardoublepage


\renewcommand{\thefootnote}{\arabic{footnote}}
\setcounter{footnote}{0}



\pagestyle{plain} 
\setcounter{page}{1}
\pagenumbering{arabic}


%

\section{Introduction}
In the last two decades a plethora of  new results 
in the charmonium spectra have been obtained 
in the beauty decays studies. 
A lot of the conventional and exotic charmonium 
resonances are observed such as $\chiconex$, $\chicone(4700)$ and $\ensuremath {\mathrm{P}_{c}(4312)^{+}}\xspace$ and conventional $\Ppsi_2(3823)$ state. 
The \lhcb experiment has collected high statistics  
during Run~1 and Run~2 periods that allows us to perform 
many precise measurements of the branching fractions 
of \B-~and \Bs-meson decays and searches for new decays 
and states.
The results described below are based on 
the data samples collected by the LHCb experiment 
in proton-proton (pp) collisions at the Large 
Hadron Collider  from 2011 to 2018 with centre-of-mass energies of $\sqs= 7,8$  and $13\tev$.

\section{Study of the 
\mbox{$\decay{\Bp }{ \jpsi \pip \pim \Kp \Km}$} decays}
\begin{figure*}[htb]
	\setlength{\unitlength}{1mm}
	\centering
	\begin{picture}(120,50)
	\definecolor{root8}{rgb}{0.35, 0.83, 0.33}

    \put(0,0){\includegraphics*[width=62mm,height=50mm]{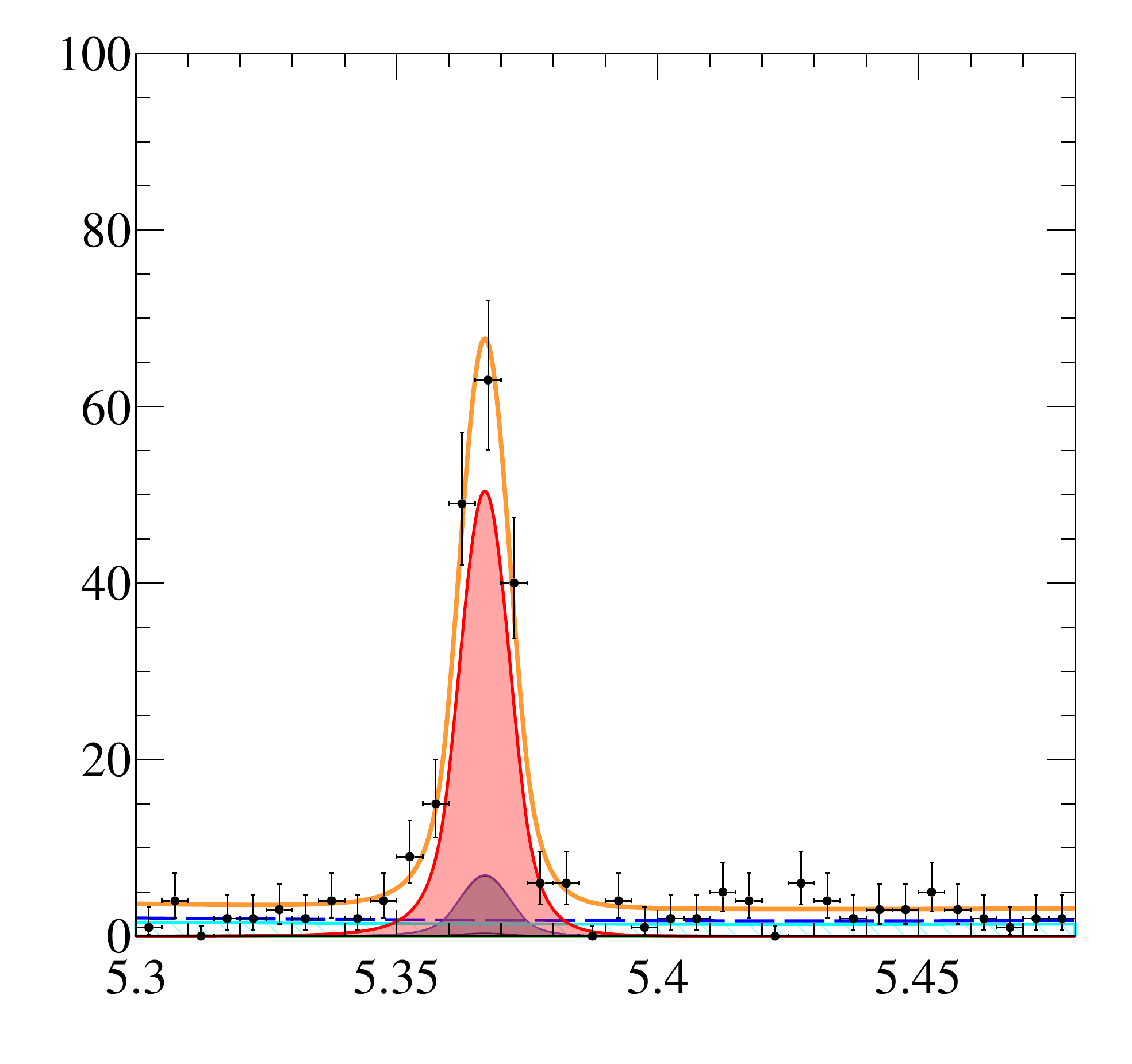}}
		\put(62,0){\includegraphics*[width=62mm,height=50mm]{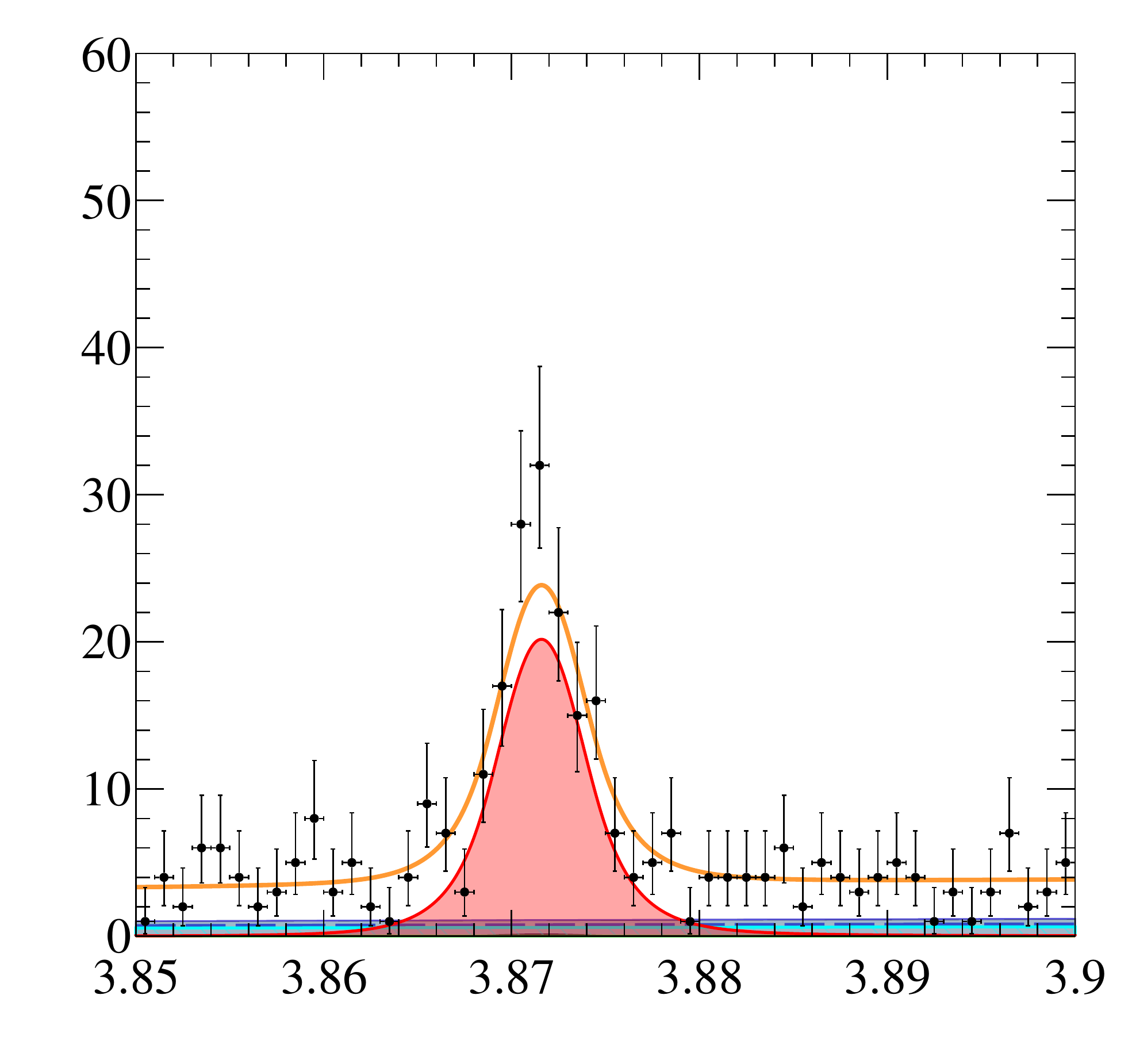}}

	\put(  -2,11){\begin{sideways}\small{Candidates/$(5\mevcc)$}\end{sideways}}
	
	\put(  62,11){\begin{sideways}\small{Candidates/$(1\mevcc)$}\end{sideways}}

	\put(20 ,0){$m_{\jpsi\pip\pim\Kp\Km}$}
	
	\put(82 ,0){$m_{\jpsi\pip\pim}$}

    \put( 46,-1){$\left[\!\gevcc\right]$}

	\put( 108,-1){$\left[\!\gevcc\right]$}

	\put( 10,38){\tiny$3.864<m_{\jpsi\pip\pim}<3.880\gevcc$}
	\put( 10,42){\tiny$1.01<m_{\Kp\Km}<1.03\gevcc$}
	\put( 71,38){\tiny$5.350<m_{\jpsi\pip\pim\Kp\Km}<5.384\gevcc$}
	\put( 71,42){\tiny$1.01<m_{\Kp\Km}<1.03\gevcc$}
	\put( 47,42){\small\lhcb}

	\put( 107,42){\small\lhcb}

	\end{picture}
		\caption {\small 
	Distributions of 
	the~(left)\,\jpsi\pip\pim\Kp \Km and 
	(right)\,\jpsi \pip \pim mass
	for selected \mbox{$\decay{\Bs}{ \chicone(3872)  \Pphi}$}~candidates\,(points with error bars)~\cite{LHCb-PAPER-2020-035}. The red filled area corresponds to the \mbox{$\decay{\Bs}{  \chicone(3872)  \Pphi}$} signal.  The orange line is the total fit.  }
	\label{fig:signal_fit3}
\end{figure*}

Candidates of  
the~\mbox{$\decay{\Bs}{ \jpsi \pip \pim  \Kp \Km}$}~decays 
are reconstructed via $\decay{\jpsi}{\mun \mup } $ and selected using
based on kinematics, particle identification 
and topology~\cite{LHCb-PAPER-2020-035}.
The yields of 
\mbox{$\decay{\Bs}{ \jpsi \pip \pim \Kp \Km}$}  decays 
via the~\mbox{$\decay{\Bs}{\psitwos \Pphi}$} and 
\mbox{$\decay{\Bs}{ \chicone(3872) \Pphi}$}
and \mbox{$\decay{\Bs}{ \jpsi\Kstarz\Kstarzb}$} chains
are determined using three-dimensional unbinned 
extended maximum-likelihood fits.
The observed signal yield for 
the \mbox{$\decay{\Bs}{ \chicone(3872)\Pphi}$} 
decays is  $154\pm 15$  which 
corresponds to a statistical significance 
more than 10 standard deviations.  
The fit to the mass distribution for the 
signal channel is shown in figure~\ref{fig:signal_fit3}. 

In addition, the decays \mbox{$\decay{\Bs}{  \chicone(3872)\Kp \Km}$}  
where the $\Kp \Km$ pair does not originate from
a $\Pphi$ meson, is studied using 
a two-dimensional unbinned extended maximum-likelihood fit which
is performed to corresponding mass distributions. 
\begin{figure}[htb]
	\setlength{\unitlength}{1mm}
	\centering
	\begin{picture}(120,75)
	\definecolor{root8}{rgb}{0.35, 0.83, 0.33}

    \put(20,00){\includegraphics*[height=75mm]{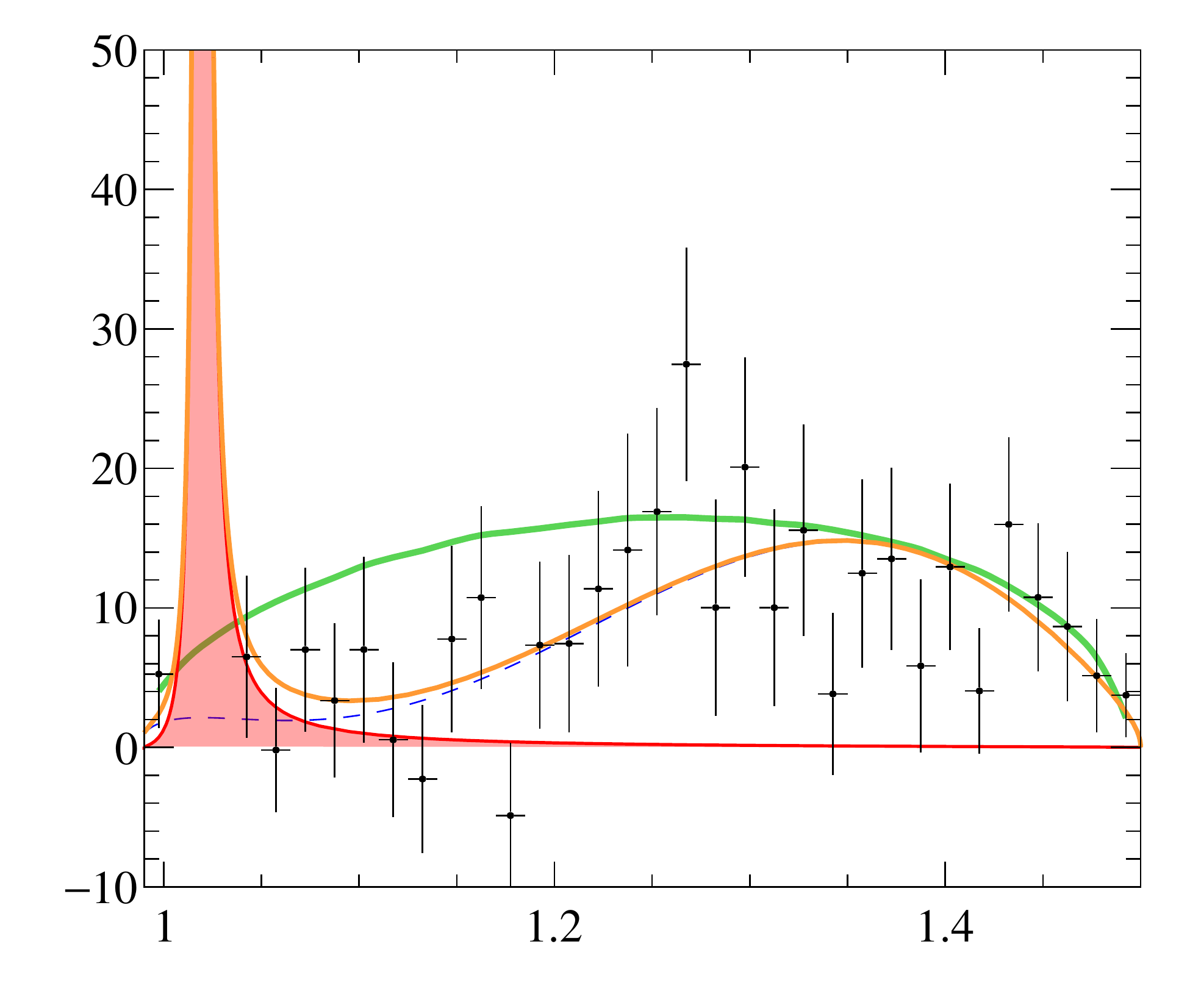}}

	\put( 18,40){\begin{sideways}Yields/$(15\mevcc)$\end{sideways}}

	\put(62,-1){\Large $m_{\Kp\Km}$}

	\put(86,-1){\Large $\left[\!\gevcc\right]$}

   \put(40,65) {\begin{tikzpicture}[x=1mm,y=1mm]\filldraw[fill=red!35!white,draw=red,thick]  (0,0) rectangle (8,3);\end{tikzpicture} }
	\put(40,62.5){\color[RGB]{85,83,246}     {\hdashrule[0.0ex][x]{8mm}{1.0pt}{1.0mm 0.5mm} } }
	\put(40,58){\color[RGB]{255,153,51} {\rule{8mm}{1.0pt}}}
	\put(50,66){\scriptsize $\decay{\Bs}{\chicone(3872)\Pphi}$}
	\put( 50,62){\scriptsize \decay{\Bs}{\chicone(3872)\Kp \Km}}
	\put( 50,58){\scriptsize total}

	\put( 92,64){\small\lhcb}

	\end{picture}
	\caption {\small 
	 Background-subtracted $\Kp\Km$~mass distribution 
	for selected \mbox{$\decay{\Bs}{  \chicone(3872)  \Kp \Km}$}~candidates\,(points with error bars)~\cite{LHCb-PAPER-2020-035}. The orange line is the total fit. }
	\label{fig:signal_fit5}
\end{figure}
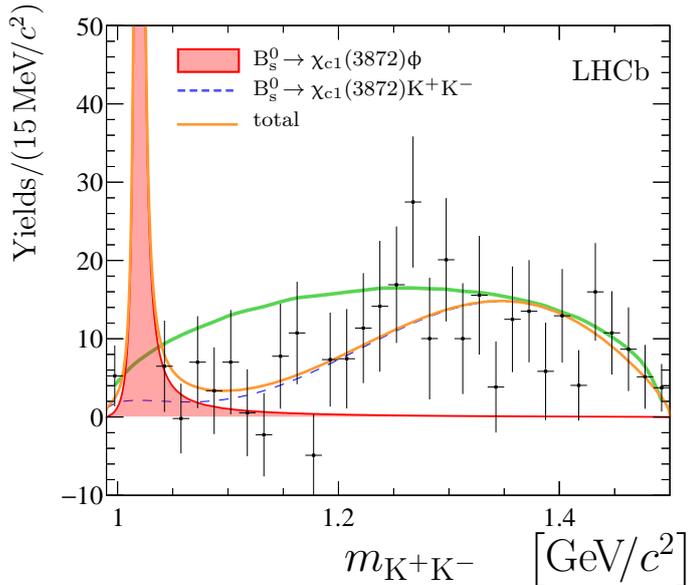
The observed yield of 
signal decays is \mbox{$378 \pm 33$},
that is significantly larger than 
the~yield of the~\mbox{$\decay{\Bs}{ \chicone(3872)\Pphi}$}~decays, 
indicating a significant \mbox{$\decay{\Bs}{  \chicone(3872)\Kp \Km}$}~contribution. 
A~narrow $\Pphi$ component can be separated from the non-$\Pphi$ components
using an unbinned maximum-likelihood fit to the background-subtracted and
efficiency-corrected $\Kp \Km$ mass distribution. 
The~fraction of
the~\mbox{$\decay{\Bs}{  \chicone(3872)\Kp \Km}$} signal component is found 
to be $(38.9 \pm 4.9)\%$.
Using the obtained signal yields and fractions 
for described channels and corresponding efficiency 
ratios the following branching fractions are calculated:
\begingroup
\allowdisplaybreaks
\begin{eqnarray*}
\dfrac{ \BRN_{\decay{\Bs}{  \chicone(3872)  \Pphi}} \times \BRN_{ \decay{\chicone(3872) } { \jpsi \pip \pim }}}
{\BRN_{\decay{\Bs}{  \psitwos\Pphi }} \times \BRN_{\decay{\psitwos }{ \jpsi \pip \pim }}}
& = & (2.42 \pm 0.23  \pm 0.07) \times 10^{-2}\,,\\
\dfrac{\BRN_{\decay{\Bs}{ \jpsi\Kstarz\Kstarzb}} 
 \times \BRN_{\decay{\Kstarz} { \Kp \pim }}^2}{\BRN_{\decay{\Bs}{ \psitwos \Pphi}} \times 
 \BRN_{ \decay{\psitwos }{ \jpsi \pip \pim} }
 \times \BRN_{\decay{\Pphi}{  \Kp\Km} }} & =  & 1.22 \pm 0.03 \pm 0.04\,, \\
\dfrac{ \BRN_{\decay{\Bs}{ \chicone(3872)  (\Kp \Km)_{\text{non-}\Pphi}} }}
{\BRN_{\decay{\Bs}{ \chicone(3872) \Pphi} } 
\times \BRN_{ \decay{\Pphi}{  \Kp\Km}  }}  & =  &  1.57 \pm 0.32  \pm 0.12 \,,
\end{eqnarray*}
\endgroup
where the~first uncertainty is statistical and the second is systematic. 
The~result for $\decay{\Bs}{  \chicone(3872)  \Pphi}$~decay  
is found to be in a good  agreement with  the result  
by the~CMS collaboration~\cite{Sirunyan:2020qir} but is more precise.

Four tetraquark candidates have been observed by
the LHCb collaboration using an amplitude 
analysis of  the~$\decay{\Bp}{ \jpsi \Pphi \Kp}$  decays~\cite{LHCb-PAPER-2016-019,LHCb-PAPER-2016-018}.
A~search of the~exotic states in 
the~$\jpsi\Pphi$ spectrum is performed 
using the~\mbox{$\decay{\Bs}{ \jpsi \pip \pim \Pphi}$} decays. 
The \mbox{$\decay{\Bs}{\jpsi \pip \pim \Pphi}$} candidates
are determined with two-di\-men\-sional  
unbinned extended 
maximum-likelihood
fit to the~$\jpsi \pip \pim \Kp \Km$ and $\Kp \Km$ mass distributions. 

\begin{figure}[htb]
	\setlength{\unitlength}{1mm}
	\centering
	\begin{picture}(120,75)
	\definecolor{root8}{rgb}{0.35, 0.83, 0.33}
		\put(20,00){\includegraphics*[height=75mm]{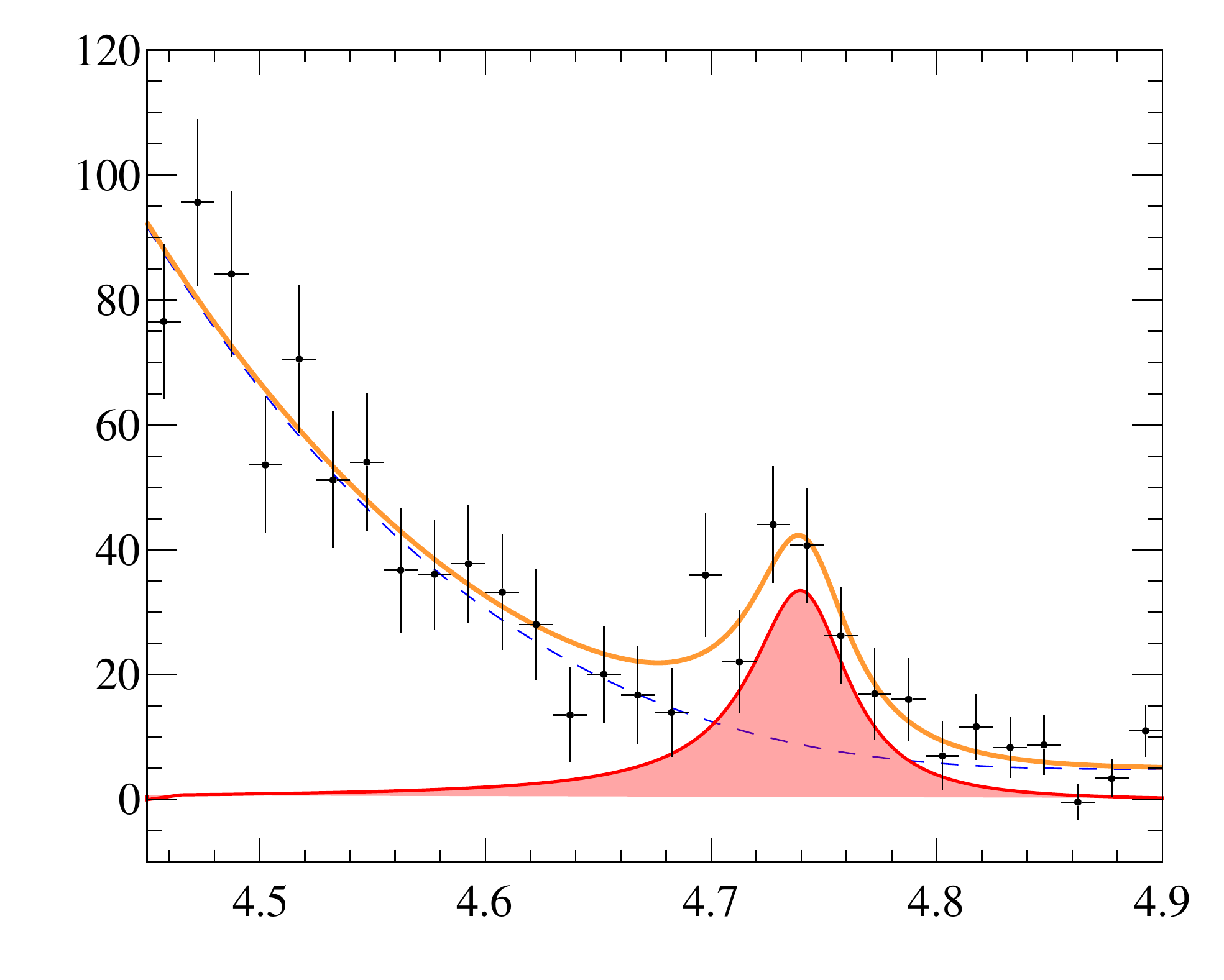}}
    \put( 64,-1){\Large$m_{\jpsi\Pphi}$}
	\put( 90,-1){\Large$\left[\!\gevcc\right]$}

    \put(18,40){\begin{sideways}Yields/$(15\mevcc)$\end{sideways}}
	\put(95,64){\small\lhcb} 

   \put(60,58){\begin{tikzpicture}[x=1mm,y=1mm]\filldraw[fill=red!35!white,draw=red,thick]  (0,0) rectangle (8,3);\end{tikzpicture} }
	\put(60,52){\color[RGB]{85,83,246}     {\hdashrule[0.0ex][x]{8mm}{1.0pt}{1.0mm 0.5mm} } }
	\put(60,47){\color[RGB]{255,153,51} {\rule{8mm}{1.0pt}}}
	\put(70,58){ \small $\decay{\Bs}{\PX(4740)\pip\pim}$}
	\put( 70,52){ \small \decay{\Bs}{\jpsi \pip \pim \Pphi}}
	\put( 70,47){ \small total}

		\end{picture}
		\caption{\small Background-subtracted $\jpsi\Pphi$~mass distribution
for the~selected \mbox{$\decay{\Bs}{  \jpsi \pip \pim \Pphi}$}~signal candidates\, (points  with error bars)~\cite{LHCb-PAPER-2020-035}.  
 The red filled area corresponds to  the \mbox{$\decay{\Bs}{  \PX(4740) \pip \pim}$} signal. 
 The orange line is the total fit.
}
		\label{fig:signal_fit6}
\end{figure}
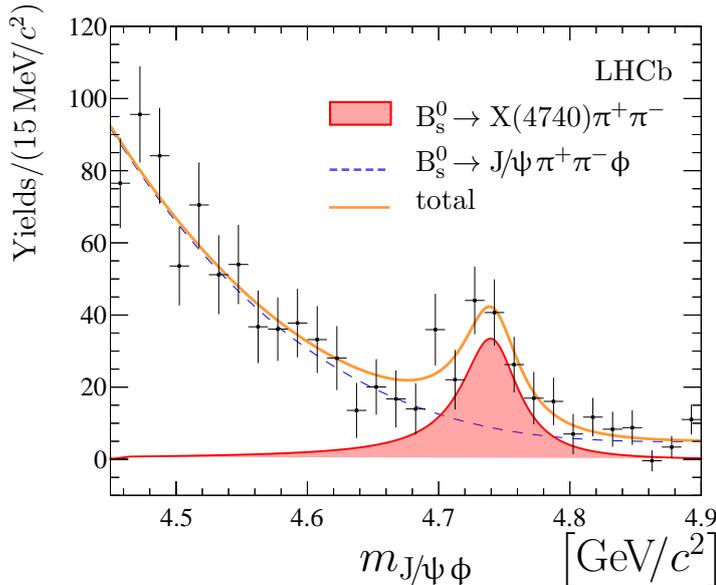

The back\-ground-sub\-tracted $\jpsi\Pphi$ mass spectrum 
of  \mbox{ $\decay{\Bs}{ \jpsi \pip \pim \Pphi}$} candidates are 
shown in figure~\ref{fig:signal_fit6}. 
It shows a~prominent structure at a mass around $4.74~\gevcc$. 
Since the regions of $\psitwos$ and $\chicone(3872)$ resonance masses are vetoed and  no sizeable contributions from decays via other narrow charmonium states 
are observed in the background-subtracted $\jpsi \pip \pim$ mass spectrum,  this structure cannot be explained by cross-feed from the $\jpsi \pip \pim$ mass spectrum.
Moreover no such structure
is seen in non-$\Pphi$  region of the $\Kp \Km$ mass.
However the~$\Pphi \pip \pim$ spectrum exhibits significant deviations from 
the phase-space distribution, indicating possible presence of excited $\Pphi$ states, 
referred to as $\Pphi^{*}$ states hereafter. 
The decays \mbox{$\decay{\Bs}{\jpsi \Pphi^{*}}$} via 
intermediate $\Pphi$(1680), $\Pphi$(1850) or $\Pphi$(2170) 
states~\cite{PDG2020} 
are studied using simulated samples and no peaking structures 
are observed. 
Under the~assumption that the~observed structure, referred to as X(4740) hereafter, 
has a~resonant nature, its mass and width are determined through 
an unbinned extended maximum-likelihood fit.
The~fit result is superimposed in figure~\ref{fig:signal_fit6}. 
The~obtained signal yield is $175\pm 39$  
and corresponds to a~statistical significance 
above 5.3~standard deviations.
The~mass and width for the $\PX(4740)$~state are found to be
\begingroup
\allowdisplaybreaks
\begin{eqnarray*}
m_{{\PX(4740)}}&= & 4741 \pm \phantom{0}6 \pm \phantom{0}6 \gevcc\,,\\ 
\Gamma_{\PX(4740)}&= &\phantom{00}53 \pm 15 \pm 11 \mev.
\end{eqnarray*}
\endgroup
The observed parameters qualitatively agree 
 with those of the $\chicone(4700)$ state observed 
 by the LHCb collaboration in references~\cite{LHCb-PAPER-2016-019,LHCb-PAPER-2016-018}.
The~obtained mass also agrees with 
the~one expected for the~$2^{++}$ \cquark\squark\cquarkbar\squarkbar tetraquark state~\cite{Ebert:2008kb}.
 
The~$\Bs$ decays to the~$\psitwos\Kp \Km$ final states characterize 
the~relatively small energy release allowing precise measurement of 
the~\Bs~meson mass. 
The~mass of the~$\Bs$  meson is determined from 
an~unbinned extended maximum-likelihood
fit to the $\psitwos\Kp \Km$ mass distribution. 
The improvement in the \Bs~mass resolution and 
significant decrease of  the~systematic uncertainties 
is achieved 
by imposing a constraint on the reconstructed 
mass of the  $\jpsi\pip\pim$~system 
to the~known  \psitwos~meson mass~\cite{PDG2020}. 
 The~measured value of the~\Bs~meson mass is found to be
 \begin{linenomath}
 \begin{equation*}
 m_{\Bs}=5366.98 \pm 0.07 \pm 0.13 \mevcc\,,     
 \end{equation*}
 \end{linenomath}
 that is the most precise single measurement of this quantity.

\section{Study of the 
\mbox{$\decay{\Bp }{ \jpsi \pip \pim \Kp}$} decays}
The search of the spin\nobreakdash-2 
component of the $D$\nobreakdash-wave 
charmonium triplet, the $\Ppsi_2(3823)$ state, is performed 
with~\mbox{$\Bp \to \jpsi \pip \pim \Kp$} decays~\cite{LHCb-PAPER-2020-009,Pereima:2745798}.
To extract the \Bp candidates, a multivariate classifier 
algorithm based on a decision tree with gradient boosting is applied. 
For  signal yield determinations of  the   
$\decay{\Bp }{ (\decay{\psitwos}{ \jpsi \pip \pim} )\Kp}$, 
$\decay{\Bp }{ (\decay{\chiconex }{ \jpsi \pip \pim} )\Kp}$ and 
$\decay{\Bp }{( \decay{\Ppsi_2(3823) }{ \jpsi \pip \pim}) \Kp}$,   
a simultaneous unbinned extended maximum-likelihood fit 
to the  $m_{\jpsi \pip \pim \Kp}$ and  $m_{\jpsi \pip \pim}$ variables 
is performed.
The signal yield for 
the $\decay{\Bp }{ \Ppsi_2(3823)  \Kp}$ decays 
is determined to be $137\pm 26$ 
which correspond to statistical significance above 
5.1~standard deviations.
Large signal yield for the~$\decay{\Bp }{ \psitwos \Kp}$~signal,
$4230 \pm 70$,  
allows for the~precise measurement 
of the~mass and width of the~$\chiconex$ state. 
For the~first time the~non-zero Breit\nobreakdash--Wigner width 
is observed for the $\chiconex$ state with significance more 
than 5~standard deviations and its measured value is:
\begin{equation*}
\Gamma_{\chicone(3872)}=0.96^{+0.19}_{-0.18} \pm 0.21\mev\, .
\end{equation*}
The~upper limit for the~Breit\nobreakdash--Wigner 
width of  $\Ppsi_{2}(3823)$ is 
improved and its value  is set 
to be $\Gamma_{\Ppsi_{2}(3823)}<5.2\,(6.6)\mev,$
for $90\,(95)\%$~C.L.
The~mass splitting between the states are found to be 
\begingroup
\allowdisplaybreaks
\begin{eqnarray*}
\Pdelta m^{\chicone(3872)}_{{\Ppsi_{2}(3823)}} & = &  
  \phantom{0}47.50\pm 0.53 \pm 0.13 \mevcc\,,\\ 
\Pdelta m^{{\Ppsi_{2}(3823)}}_{{\psitwos}} & = & 
137.98\pm 0.53 \pm 0.14 \mevcc\,, \\ 
\Pdelta m^{{\Ppsi_{2}(3823)}}_{{\psitwos}} & = & 
185.49\pm 0.06 \pm 0.03 \mevcc\,,
\end{eqnarray*}
\endgroup
The~results  Breit\nobreakdash--Wigner
mass of the~\chiconex~state are in good agreement
with an independent analysis of 
inclusive $\decay{\bquark}{\chiconex\PX}$~decays~\cite{LHCb-PAPER-2020-008}.
The binding energy of the~$\chicone(3872)$ state is derived from the mass splitting  and its value is found to be 
\mbox{$\delta E=0.12\pm 0.13 \mev$}.
  It is consistent 
with zero within uncertainties, that are currently 
dominated   by the~uncertainty  
for the~neutral and charged kaon mass 
measurements~\cite{Tomaradze:2014eha,LHCb-PAPER-2013-011}.

The measured yields of 
the~\mbox{$\decay{\Bp}{\chicone(3872) \Kp}$},
\mbox{$\decay{\Bp}{\Ppsi_{2}(3823) \Kp}$} and 
\mbox{$\decay{\Bp}{\psitwos \Kp}$} signal decays
allow for a precise determination of the ratios of 
the~branching fractions:
\begingroup
\allowdisplaybreaks
\begin{eqnarray*}
\dfrac{ \BRN_{\decay{\Bp }{  \Ppsi_{2}(3823)  \Kp}} \times \BRN_{\decay{\Ppsi_{2}(3823)} {\jpsi \pip \pim }}}
{\BRN_{\decay{\Bp }{  \chicone(3872)  \Kp}} \times 
\BRN_{\decay{\chicone(3872) } { \jpsi \pip \pim}  } }
& = &     \left( 3.56 \pm 0.67 \pm 0.11 \right) \times 10^{-2} \,, 
\\
\dfrac{ \BRN_{\decay{\Bp }{ \Ppsi_{2}(3823)  \Kp }} \times 
\BRN_{ \decay{\Ppsi_{2}(3823)} {\jpsi \pip \pim }}}
{\BRN_{\decay{\Bp }{  \psitwos \Kp} } \times \BRN_{ \decay{\psitwos} {  \jpsi \pip \pim }}}
& = &  (1.31 \pm 0.25 \pm 0.04) \times 10^{-3} \,,  \\
\dfrac{ \BRN_{\decay{\Bp }{  \chicone(3872)  \Kp }} \times \BRN_{\decay{\chicone(3872) } { \jpsi \pip \pim} }}
{\BRN_{\decay{\Bp }{ \psitwos\Kp} } \times \BRN_{ \decay{\psitwos} { \jpsi \pip \pim} }} 
& = & 
(3.69 \pm 0.07  \pm 0.06) \times 10^{-2} \,.
\end{eqnarray*}
\endgroup

 \section{Conclusion}
 A~study of $b$\nobreakdash-meson decays 
\mbox{$\decay{\Bp}{\jpsi \pip \pim \Kp}$} and 
\mbox{$\decay{\Bs}{\jpsi\pip\pim\Kp\Km}$} 
is made using the Run 1 and Run 2 data, 
collected with the~LHCb detector~\cite{LHCb-PAPER-2020-035,LHCb-PAPER-2020-009}.
The~reported results include 
the~ observation of the non-zero width of 
the~$\chicone(3872)$~state; 
the most precise measurement of 
the~masses of the~$\chicone(3872)$ and $\Ppsi_2(3823)$~states; 
the~most precise measurement of several ratios of  
branching fractions of  the~\Bu and \Bs~mesons decays; 
the~most precise single measurement of the \Bs~meson mass
and the~observation of a~new structure, denoted as the~$\PX(4740)$~state,
in the~$\jpsi\Pphi$~mass spectrum.

\addcontentsline{toc}{section}{References}
\bibliographystyle{LHCb}
\bibliography{main,standard,LHCb-PAPER,LHCb-CONF,LHCb-DP,LHCb-TDR}
 


\end{document}